\documentclass[aps,prl,twocolumn,superscriptaddress,showpacs,preprintnumbers,amsmath,amssymb,floatfix]{revtex4}

\usepackage{graphicx}
\usepackage{dcolumn}
\usepackage{bm}
\usepackage{url} 
\usepackage{amsmath} 
\usepackage{amssymb} 

\begin{document}

\title{\quad\\[1.0cm] Observation of 
$D^+ \rightarrow K^{+} \eta^{(\prime)}$
and Search for $CP$ Violation in
$D^+ \rightarrow \pi^{+} \eta^{(\prime)}$
Decays
}

\affiliation{Budker Institute of Nuclear Physics SB RAS and Novosibirsk State University, Novosibirsk 630090}
\affiliation{Faculty of Mathematics and Physics, Charles University, Prague}
\affiliation{University of Cincinnati, Cincinnati, Ohio 45221}
\affiliation{Justus-Liebig-Universit\"at Gie\ss{}en, Gie\ss{}en}
\affiliation{Gifu University, Gifu}
\affiliation{Gyeongsang National University, Chinju}
\affiliation{Hanyang University, Seoul}
\affiliation{University of Hawaii, Honolulu, Hawaii 96822}
\affiliation{High Energy Accelerator Research Organization (KEK), Tsukuba}
\affiliation{Hiroshima Institute of Technology, Hiroshima}
\affiliation{Indian Institute of Technology Guwahati, Guwahati}
\affiliation{Indian Institute of Technology Madras, Madras}
\affiliation{Institute of High Energy Physics, Chinese Academy of Sciences, Beijing}
\affiliation{Institute of High Energy Physics, Vienna}
\affiliation{Institute of High Energy Physics, Protvino}
\affiliation{Institute for Theoretical and Experimental Physics, Moscow}
\affiliation{J. Stefan Institute, Ljubljana}
\affiliation{Kanagawa University, Yokohama}
\affiliation{Institut f\"ur Experimentelle Kernphysik, Karlsruher Institut f\"ur Technologie, Karlsruhe}
\affiliation{Korea Institute of Science and Technology Information, Daejeon}
\affiliation{Korea University, Seoul}
\affiliation{Kyungpook National University, Taegu}
\affiliation{\'Ecole Polytechnique F\'ed\'erale de Lausanne (EPFL), Lausanne}
\affiliation{Faculty of Mathematics and Physics, University of Ljubljana, Ljubljana}
\affiliation{University of Maribor, Maribor}
\affiliation{Max-Planck-Institut f\"ur Physik, M\"unchen}
\affiliation{University of Melbourne, School of Physics, Victoria 3010}
\affiliation{Nagoya University, Nagoya}
\affiliation{Nara Women's University, Nara}
\affiliation{National Central University, Chung-li}
\affiliation{National United University, Miao Li}
\affiliation{Department of Physics, National Taiwan University, Taipei}
\affiliation{H. Niewodniczanski Institute of Nuclear Physics, Krakow}
\affiliation{Nippon Dental University, Niigata}
\affiliation{Niigata University, Niigata}
\affiliation{University of Nova Gorica, Nova Gorica}
\affiliation{Osaka City University, Osaka}
\affiliation{Pacific Northwest National Laboratory, Richland, Washington 99352}
\affiliation{Panjab University, Chandigarh}
\affiliation{Research Center for Nuclear Physics, Osaka}
\affiliation{Saga University, Saga}
\affiliation{University of Science and Technology of China, Hefei}
\affiliation{Seoul National University, Seoul}
\affiliation{Sungkyunkwan University, Suwon}
\affiliation{School of Physics, University of Sydney, NSW 2006}
\affiliation{Tata Institute of Fundamental Research, Mumbai}
\affiliation{Excellence Cluster Universe, Technische Universit\"at M\"unchen, Garching}
\affiliation{Toho University, Funabashi}
\affiliation{Tohoku Gakuin University, Tagajo}
\affiliation{Tohoku University, Sendai}
\affiliation{Department of Physics, University of Tokyo, Tokyo}
\affiliation{Tokyo Institute of Technology, Tokyo}
\affiliation{Tokyo Metropolitan University, Tokyo}
\affiliation{Tokyo University of Agriculture and Technology, Tokyo}
\affiliation{CNP, Virginia Polytechnic Institute and State University, Blacksburg, Virginia 24061}
\affiliation{Yonsei University, Seoul}
  \author{E.~Won}\affiliation{Korea University, Seoul} 
  \author{B.~R.~Ko}\affiliation{Korea University, Seoul} 
  \author{I.~Adachi}\affiliation{High Energy Accelerator Research Organization (KEK), Tsukuba} 
  \author{H.~Aihara}\affiliation{Department of Physics, University of Tokyo, Tokyo} 
  \author{K.~Arinstein}\affiliation{Budker Institute of Nuclear Physics SB RAS and Novosibirsk State University, Novosibirsk 630090} 
  \author{D.~M.~Asner}\affiliation{Pacific Northwest National Laboratory, Richland, Washington 99352} 
  \author{T.~Aushev}\affiliation{Institute for Theoretical and Experimental Physics, Moscow} 
  \author{A.~M.~Bakich}\affiliation{School of Physics, University of Sydney, NSW 2006} 
  \author{E.~Barberio}\affiliation{University of Melbourne, School of Physics, Victoria 3010} 
  \author{A.~Bay}\affiliation{\'Ecole Polytechnique F\'ed\'erale de Lausanne (EPFL), Lausanne} 
% ...
\collaboration{The Belle Collaboration}

\begin{abstract}

 We report the first observation of the doubly Cabibbo-suppressed decays
$D^+ \rightarrow K^+ \eta^{(\prime)}$ 
using a 791 fb$^{-1}$ 
data sample collected with the Belle detector
at the KEKB asymmetric-energy $e^+e^-$ collider.
The ratio of the branching fractions of doubly Cabibbo-suppressed relative 
to singly Cabibbo-suppressed $D^+ \rightarrow \pi^+ \eta^{(\prime)}$ decays are
$\mathcal{B}(D^+ \rightarrow K^+ \eta)$/$\mathcal{B}(D^+ \rightarrow \pi^+ \eta)$
= (3.06 $\pm$ 0.43 $\pm$ 0.14)\%
and
$\mathcal{B}(D^+ \rightarrow K^+ \eta')$/$\mathcal{B}(D^+ \rightarrow \pi^+ \eta')$
= (3.77 $\pm$ 0.39 $\pm$ 0.10)\%.
From these, we find that the relative final-state phase difference between the 
tree and annihilation amplitudes in $D^+$ decays,
$\delta_{TA}$, is (72 $\pm$ 9)$^\circ$ or
(288 $\pm$ 9)$^\circ$.
We also report the most precise measurements of $CP$ asymmetries to date:
$A_{CP}^{D^+ \rightarrow \pi^+ \eta}$
=
($+$1.74 $\pm$ 1.13 $\pm$ 0.19)\% and  
$A_{CP}^{D^+ \rightarrow \pi^+ \eta'}$
=
($-$0.12 $\pm$ 1.12 $\pm$ 0.17)\%.
\end{abstract}


\maketitle

{\renewcommand{\thefootnote}{\fnsymbol{footnote}}}
\setcounter{footnote}{0}

%
%

 Decays of charmed mesons play an important role in understanding the sources
of SU(3) flavor symmetry breaking structure~\cite{ref:rosner,ref:cheng} and 
can also be sensitive probes of the violation of the combined charge-conjugation
and parity symmetries ($CP$) produced by the irreducible complex phase in the 
Cabibbo-Kobayashi-Maskawa flavor-mixing matrix~\cite{ref:ckm} 
in the standard model (SM). This SU(3) flavor symmetry structure 
is not well studied in $D^+$ meson decays 
into two-body final states with an $\eta^{(\prime)}$, since they are all
Cabibbo-suppressed decays. 
Examples of two-body decays with an $\eta^{(\prime)}$ in the final state are the doubly
Cabibbo-suppressed (DCS) decays  $D^+\rightarrow K^+ \eta^{(\prime)}$
and the singly Cabibbo-suppressed (SCS) 
decays $D^+ \rightarrow \pi^+ \eta^{(\prime)}$.
The DCS decays $D^+ \rightarrow K^+ \eta^{(\prime)}$ 
have not yet been observed. The observation of 
such modes is not only intrinsically important to illuminate 
the meson decay process but also there is general interest in the experimental
technique of measuring an extremely rare decay 
processes with neutral particles.  
Observation of $D^+\rightarrow K^+ \eta^{(\prime)}$ would complete
the picture of DCS decays for 
$D^+$ mesons decaying to pairs of light pseudoscalar mesons.

 In this Letter, we report the first observation of
$D^+ \rightarrow K^+ \eta^{(\prime)}$ decays.
The DCS decays $D^+\rightarrow K^+ \eta^{(\prime)}$ together with $D^+\rightarrow K^+ \pi^0$
can be used to measure 
the relative phase difference between
the tree and annihilation amplitudes ($\delta_{TA}$), which is an
important piece of 
information relevant to final-state interactions in $D$ meson
decays.
Note that experimentally one is able to determine only 
the tree and annihilation amplitudes
and the relative phase difference between them since all decays 
involving $K^0$ will be overwhelmed by Cabibbo-favored decays involving
a $\bar{K}^0$, with no way to distinguish between them 
because one detects only a $K^0_S$~\cite{ref:dcsd_th}.
In addition, the most sensitive search for $CP$ violation
in $D^+ \rightarrow \pi^+ \eta^{(\prime)}$ decays is reported. 
Observation of $CP$ violation in 
$D^+ \rightarrow \pi^+ \eta^{(\prime)}$ decays 
with current experimental sensitivity
would represent strong evidence for processes involving physics 
beyond the SM~\cite{ref:new}.

 The data used in this analysis were recorded at or near the $\Upsilon(4S)$ 
resonance
with the Belle detector~\cite{ref:belle} at the $e^+e^-$ 
asymmetric-energy collider KEKB~\cite{ref:kekb}. The sample corresponds to an
integrated luminosity of 791 fb$^{-1}$.

%
%

\begin{figure}[htbp]
\includegraphics[width=0.50\textwidth]{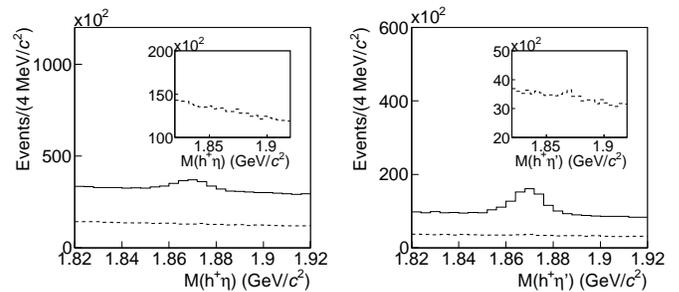}
\caption{
The invariant mass distributions of $h^+ \eta$ 
($h^+ \eta'$) in the left (right) plot after the initial selection.
The solid histograms show $\pi^+ \eta^{(\prime)}$ while the dashed histograms
show  $K^+ \eta^{(\prime)}$ final states.
The two inset histograms are $K^+ \eta^{(\prime)}$ decays with enlarged
vertical scales. 
}
\label{fig:standard}
\end{figure}

 We apply the same charged track selection criteria that were used in Ref.~\cite{ref:won_br}.
Charged kaons and pions are identified by requiring the ratio of particle identification (PID)
likelihoods~\cite{ref:won_br} to be greater or less than 0.6, respectively.
For kaons (pions) used in this analysis, the efficiencies 
and misidentification probabilities
are approximately 87\% (88\%) and 9\% (10\%), respectively.
For the reconstruction of the $\eta$ meson in 
the $D^+ \rightarrow h^+ \eta$ decay,
where $h^+$ refers to either $\pi^+$ or $K^+$, we use 
the $\eta \rightarrow \pi^+\pi^- \pi^0$ mode instead of the frequently 
used $\eta \rightarrow
\gamma \gamma$ ($\eta_{\gamma\gamma}$) mode since our event selection will include stringent 
requirements on the vertex formed from charged tracks in the $\eta$ decay.
We find that the $\eta \rightarrow \gamma \gamma$ mode has a small signal
to background ratio and poor $\eta$ invariant
mass resolution that prohibit the final signal extraction from our data.
To reconstruct the $\eta'$ meson
in $D^+ \rightarrow h^+ \eta^\prime$ decay, we use the 
$\eta^\prime \rightarrow \pi^+ \pi^-
\eta_{\gamma \gamma}$ decay.
The minimum energy of the $\gamma$ from the $\pi^0$ or $\eta$
is chosen to be 60 MeV for the barrel
and 100 MeV for the forward region of the calorimeter~\cite{ref:calorimeter}.
The decay vertex of the $D^+$ is formed by fitting the three charged tracks
($h^+ \pi^+ \pi^-$) to a common vertex and requiring a confidence level (C.L.)
greater than 0.1\%. 
%
%
For $\pi^0$ reconstruction in $D^+ \rightarrow h^+ \eta$,
we require the invariant mass of 
the $\gamma \gamma$ pair to be within [0.12,0.15] GeV/$c^2$
and for the $\eta$ we require the invariant mass of 
the $\pi^+ \pi^- \pi^0$ system to be within [0.538,0.558] GeV/$c^2$.
%
%
In the $D^+ \rightarrow h^+ \eta'$ mode, to reconstruct
the daughter $\eta_{\gamma \gamma}$, we require the invariant mass of 
the $\gamma \gamma$ pair to be 
within [0.50,0.58] GeV/$c^2$. Furthermore, in order to remove a 
significant $\pi^0$
contribution under the $\eta_{\gamma \gamma}$ signal peak, we 
reject $\gamma$ candidates as described in
Ref.~\cite{ref:brko_hp0}.
The $\pi^+ \pi^- \eta_{\gamma \gamma}$ invariant mass is required to be within
the range [0.945,0.970] GeV/$c^2$. 
The momenta of photons from the $\pi^0$ and the $\eta_{\gamma \gamma}$ 
combination are
recalculated with $\pi^0$ and $\eta$ mass~\cite{ref:pdg2010}
constraints, respectively.
The invariant mass distributions
of the $h^+ \eta^{(\prime)}$ system after the initial selection described above
are shown in Fig.~\ref{fig:standard} where
there is little indication of signal for either of the DCS modes. 
%
%

\begin{figure}[htbp]
\mbox{
\includegraphics[width=0.22\textwidth]{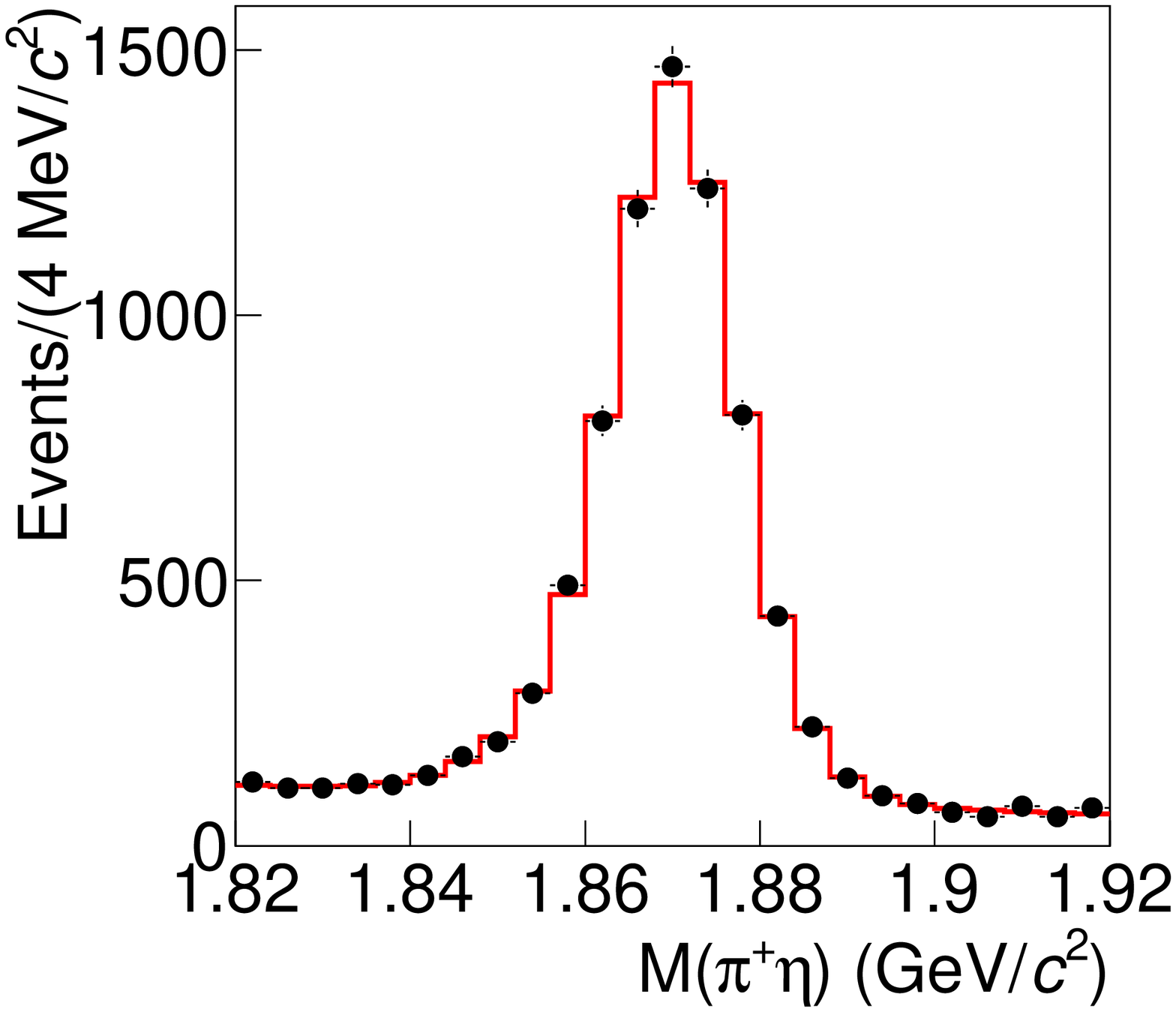} \quad
\includegraphics[width=0.22\textwidth]{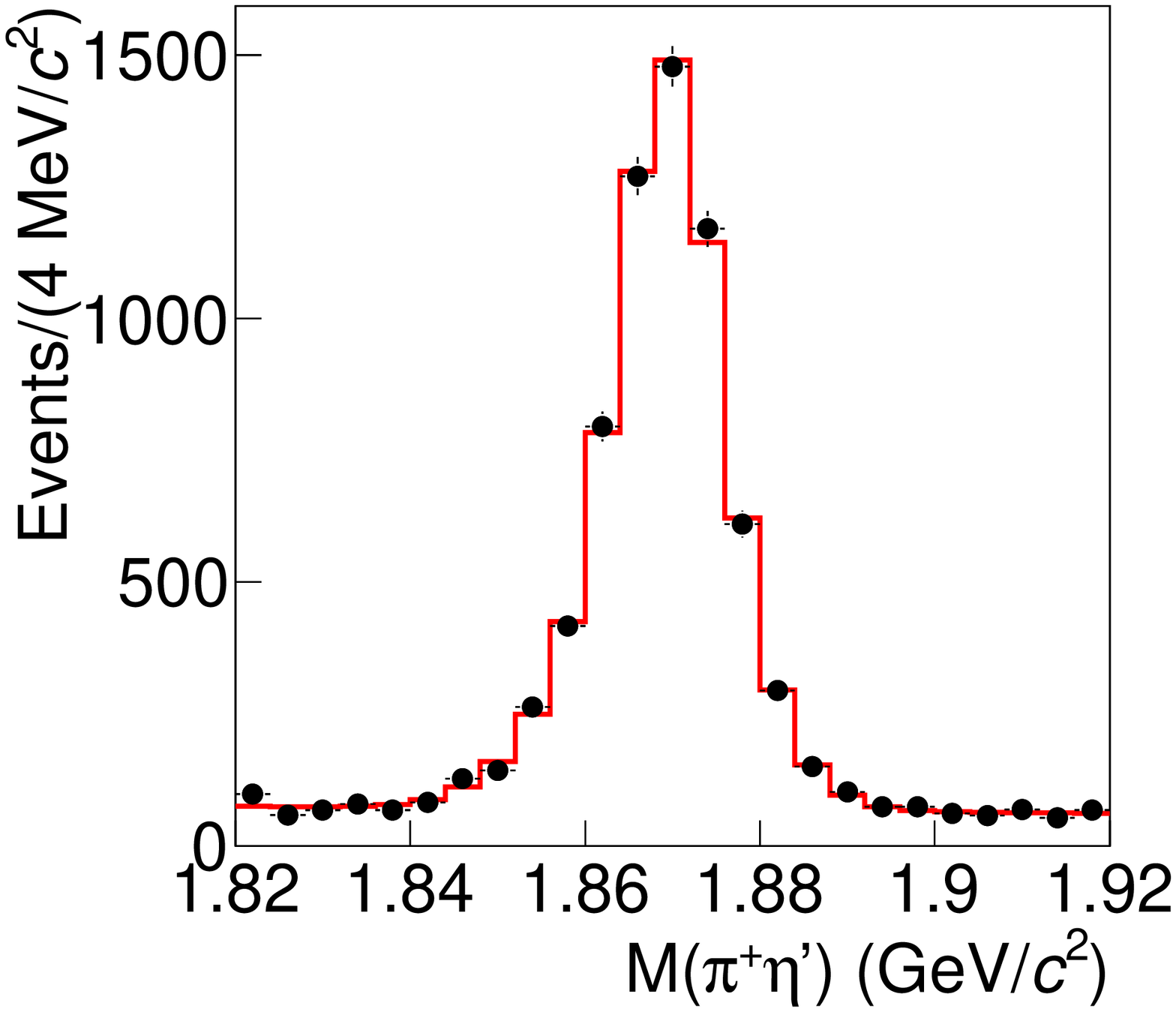}
}
\mbox{
\includegraphics[width=0.22\textwidth]{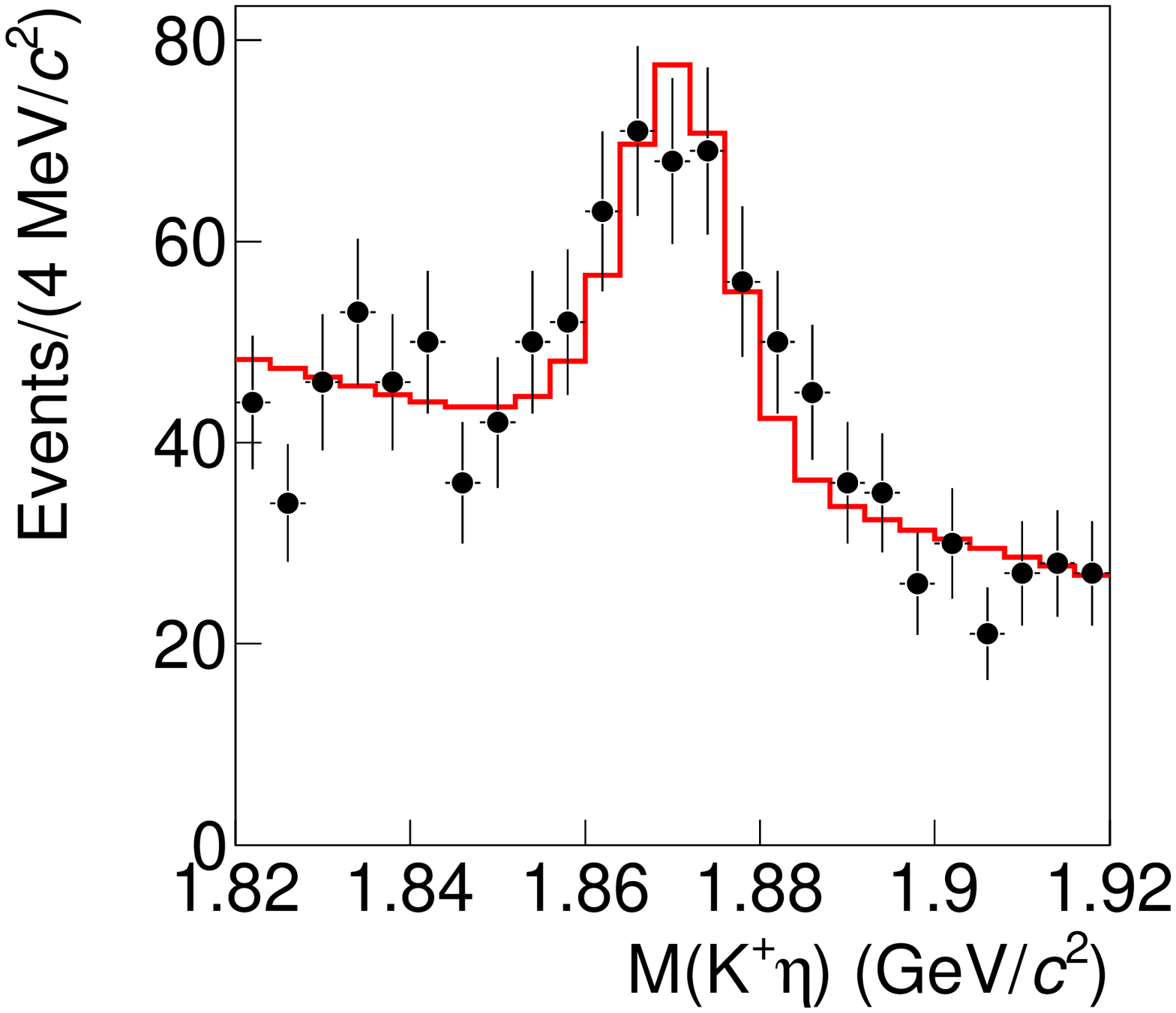} \quad
\includegraphics[width=0.22\textwidth]{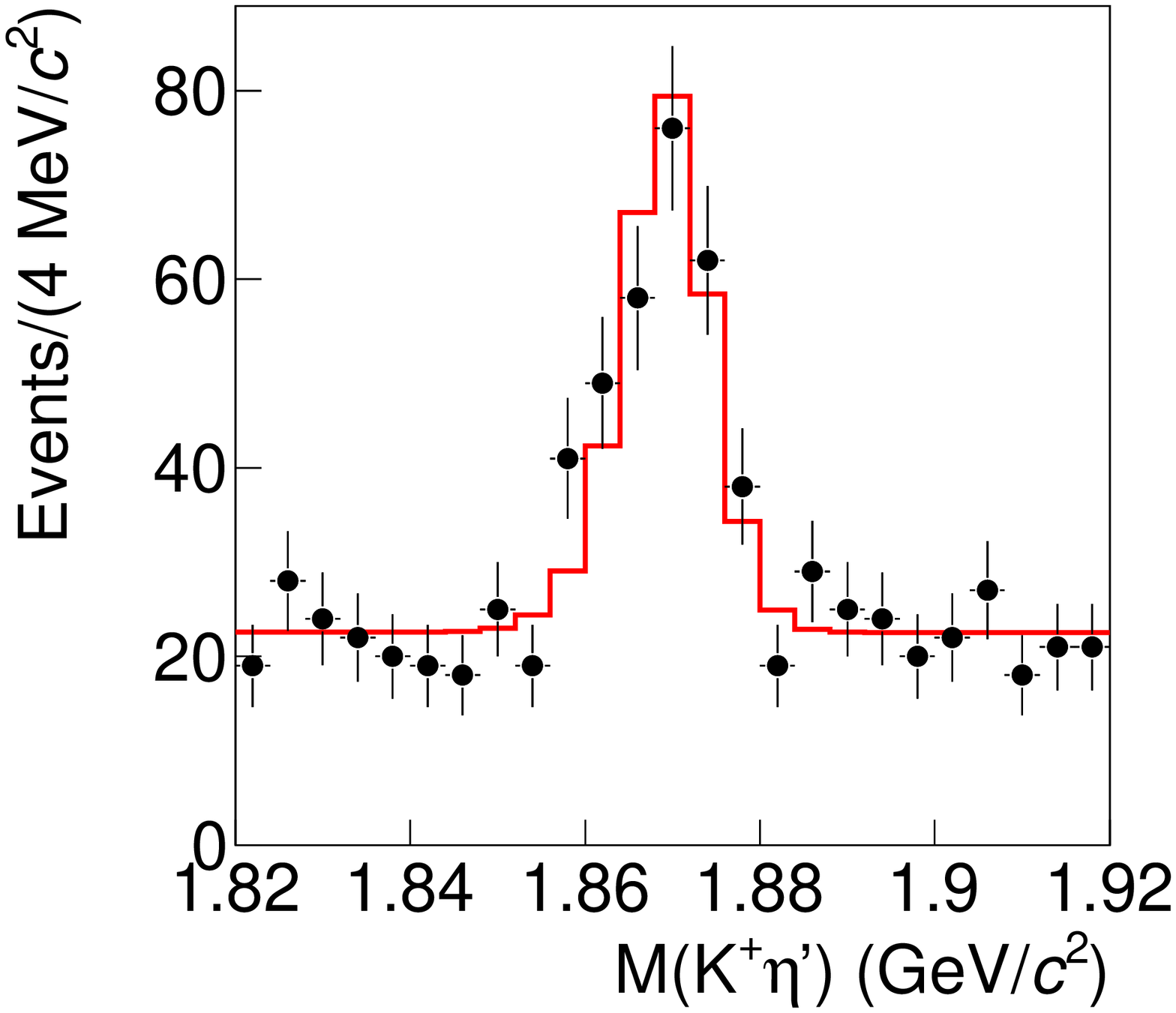}
}
\caption{
The invariant mass distributions used for the branching fraction measurements.
The top two plots are for the 
$\pi^+ \eta$ (left) and $\pi^+ \eta'$ (right) final states
while the bottom two plots are for the
$K^+ \eta$ (left) and $K^+ \eta'$ (right) final states.
Points with error bars and histograms correspond to the data and 
the fit, respectively.
}
\label{fig:dmass}
\end{figure}

%
%
 In order to search for $D^+ \rightarrow K^+ \eta^{(\prime)}$ decays, the 
following four variables are considered. The first is 
the angle ($\xi$) between the charmed meson momentum vector, as 
reconstructed from the daughter particles, and
the vector joining its production and decay vertices~\cite{ref:dcsd}. 
The second variable is 
the isolation $\chi^2$ ($\chi^2_\textrm{iso}$) 
normalized by the number of degrees of freedom (d.o.f)
for the hypothesis that the
candidate tracks forming the charmed meson arise 
from the primary vertex, where the primary vertex is 
the most probable point of intersection of the charmed meson momentum vector
and the $e^+e^-$ interaction region~\cite{ref:dcsd}. 
Because of the finite lifetime of $D^+$ mesons their daughter tracks
are not likely to be compatible with the primary vertex.
The third and the fourth variables are 
the momentum of the $\eta^{(\prime)}$ ($p_{\eta^{(\prime)}}$)
in the laboratory system, and the momentum of the
$D^+$ in the center-of-mass system ($p^*_{D^{+}}$). To optimize
the selection, we maximize $\epsilon_\textrm{sig}/\sqrt{\mathcal{N}_B}$ where $\epsilon_\textrm{sig}$ 
and $\mathcal{N}_B$
are the signal efficiency and the background yield in the invariant mass
distribution of $D^+$ candidates. A uniform grid of 10,000 points in four 
dimensions
spanned by the four kinematic variables described above is used to select 
an optimal set of selection requirements using Monte Carlo (MC) simulation 
samples~\cite{ref:MC}. Since we use MC samples, this is similar to
the importance-sampled grid search technique in Ref.~\cite{ref:grid}.
The optimal selection for the $D^+ \rightarrow K^+ \eta$ mode is found to be:
$\xi$ $<$ 5$^\circ$,
$\chi^2_\textrm{iso}$ $>$ 10,
$p_\eta$ $>$ 1 GeV/$c$, 
and 
$p^*_{D^{+}}$ $>$ 3 GeV/$c$, and for $D^+ \rightarrow K^+ \eta'$ is : 
$\xi$ $<$ 5$^\circ$,
$\chi^2_\textrm{iso}$ $>$ 5,
$p_{\eta'}$ $>$ 1.5 GeV/$c$, 
and 
$p^*_{D^{+}}$ $>$ 3 GeV/$c$. 
The same selection criteria are applied to
the normalization modes, $D^+ \rightarrow \pi^+ \eta^{(\prime)}$.
Figure ~\ref{fig:dmass} shows 
the $\pi^+ \eta^{(\prime)}$
and
$K^+ \eta^{(\prime)}$
invariant mass distributions after the final selections used for the
branching fraction measurements.
Possible structures, for example from
$D^+_s \rightarrow K^+ \pi^- \pi^+ \pi^0$ or 
$D^+_s \rightarrow K^+ K^- \pi^+ \pi^0$  
due to
particle misidentification or cross-feed between $\eta$ and $\eta^\prime$
are studied using MC samples; we find no indication of such background.

A fit is then performed for $D^+ \rightarrow \pi^+ \eta^{(\prime)}$ candidates 
and the results are shown as the 
top two
plots in Fig.~\ref{fig:dmass}.
The signal probability density function (PDF) is modeled as 
the sum of a Gaussian and a bifurcated Gaussian
while the combinatorial background is modeled as a linear background. 
The $\chi^2$/d.o.f of fits are 0.7 and 1.4, respectively. 
For fits to these DCS decays, we fix the width of the Gaussian, 
the two widths of the
bifurcated Gaussian, and then ratio of the 
normalizations of the
Gaussian and the bifurcated Gaussian to the
values obtained from the fits to the SCS modes in order to obtain stable fits. 
The fixed widths are scaled according to the difference of widths observed in 
the signal MC samples. We examine possible
systematic uncertainties due to this later. The statistical significance 
of the signal based on the log-likelihood
ratio is 9$\sigma$ and more than 10$\sigma$ 
($\sigma$ represents one standard deviation from the background-only 
hypothesis)
for 
$D^+ \rightarrow K^+ \eta$
and
$D^+ \rightarrow K^+ \eta'$, respectively; 
the corresponding invariant mass distributions
and fits
are shown in the lower panel of Fig.~\ref{fig:dmass}. The $\chi^2$/d.o.f of 
fits to the $K^+ \eta$ and
$K^+ \eta'$ final states are 0.8 and 0.9, respectively.
In order to compute the ratio of branching fractions of DCS 
modes with respect to SCS 
modes, the signal efficiencies for the selection criteria
described above are estimated with our signal MC samples.  
Table~\ref{table:eff} lists all the information used for the
branching fraction measurements.

%
%

%
%
\begin{table}
\caption{
Yields from the data and the signal efficiencies for the
branching fraction measurements. Errors are statistical only.
}
\label{table:eff} 
\begin{ruledtabular}
\begin{tabular}{lrr} 
Mode  & yield& Signal Efficiency (\%)\\
\hline
$D^{+}\rightarrow K^+ \eta $   &  166 $\pm$ ~23 & 1.35 $\pm$ 0.01 \\
$D^{+}\rightarrow K^+ \eta'$   &  180 $\pm$ ~19 & 1.20 $\pm$ 0.01 \\
$D^{+}\rightarrow \pi^+ \eta $ & 6476 $\pm$ 110 & 1.68 $\pm$ 0.02 \\
$D^{+}\rightarrow \pi^+ \eta'$ & 6023 $\pm$ ~93 & 1.59 $\pm$ 0.01 \\ 
\end{tabular}     
\end{ruledtabular}
\end{table}
The dominant  sources 
of the systematic uncertainty in the branching fraction 
measurements are the uncertainties of the parameters that are fixed in
the fits to DCS decays, and are estimated to be 
3.4\% (2.1\%) for the $\eta~(\eta')$ mode. 
These uncertainties are determined
by refitting the data with the fit parameters varied
by one standard deviation.
Other sources include the choice of
the fitting functions, estimated to be 2.7\% (1.0\%) for 
the $\eta~(\eta')$ mode, and the uncertainty in the 
PID, estimated to be 1.1\% for the both modes.
A summary of the systematic uncertainties for the ratio of branching fraction 
measurements can be found in Table~\ref{table:br_sys}.
The ratios of branching fractions are 
$\mathcal{B}(D^+ \rightarrow K^+ \eta)$/$\mathcal{B}(D^+ \rightarrow \pi^+ \eta)$
= (3.06 $\pm$ 0.43 $\pm$ 0.14)\%
and
$\mathcal{B}(D^+ \rightarrow K^+ \eta')$/$\mathcal{B}(D^+ \rightarrow \pi^+ \eta')$
= (3.77 $\pm$ 0.39 $\pm$ 0.10)\%.
We use the measurements 
of the SCS modes from 
Ref.~\cite{ref:cleo} to calculate the absolute branching fractions.
Table~\ref{table:summary} shows the comparison of our branching 
fractions
with the best present limits from Ref.~\cite{ref:cleo}. 
While the measured branching fraction for the $K^+ \eta$ mode is in agreement
with the SU(3) based
expectations~\cite{ref:rosner,ref:cheng}, 
the $K^+ \eta'$ mode is measured to be larger, by approximately three 
standard deviations. 

\begin{table}[htb]
\caption{Summary of all relative systematic uncertainties for the 
measurements of ratios of branching fractions.
}
\label{table:br_sys}
\begin{ruledtabular}
\begin{tabular}{lrr}
Source & $\sigma\Big(\frac{\mathcal{B}(D^+ \rightarrow K^+ \eta)}{\mathcal{B}(D^+ \rightarrow \pi^+ \eta )}\Big)$ (\%) 
& $\sigma\Big(\frac{\mathcal{B}(D^+\rightarrow  K^+\eta')}{\mathcal{B}(D^+ \rightarrow  \pi^+\eta')}\Big)$ (\%) 
\\
\hline
PID & 1.1 & 1.1 \\
Signal PDF & 3.4 & 2.1 \\
Fit method & 2.7 & 1.0 \\
\hline
Total & 4.5 & 2.6 \\ 
\end{tabular}
\end{ruledtabular}
\end{table}

Using the relations in Ref.~\cite{ref:dcsd_th}, which give
\begin{eqnarray}
|T|^2 &=& 3|\mathcal{A}(K^+\eta)|^2
\nonumber \\
|A|^2 &=& \frac{1}{2}\Bigg[|\mathcal{A}(K^+\pi^0)|^2
+|\mathcal{A}(K^+ \eta^\prime)|^2\Bigg] -
|\mathcal{A}(K^+\eta)|^2
\nonumber \\
\cos{\delta_{TA}} &=& \frac{1}{2|T||A|}
\Bigg[
2|\mathcal{A}(K^+\eta)|^2 + \frac{1}{2}|\mathcal{A}(K^+\eta^\prime)|^2 
\nonumber \\
&-& \frac{3}{2}|\mathcal{A}(K^+ \pi^0)|^2
\Bigg]
\end{eqnarray}
where $T$ ($A$) is the tree (annihilation) amplitude 
and $\mathcal{A}$ is the specified decay amplitude, and
from the recent branching fraction measurement of $\mathcal{B}(D^+ \rightarrow
K^+ \pi^0) = (1.72\pm 0.20) \times 10^{-4} $~\cite{ref:cleo}, we find that 
the relative final-state phase difference between the 
tree and annihilation in $D^+$ decays,
$\delta_{TA}$, is (72 $\pm$ 9)$^\circ$ or
(288 $\pm$ 9)$^\circ$.

\begin{table}[htb]
\caption{Comparison of our branching fraction results 
to 
the present best upper limit (90\% C.L.) from Ref.~\cite{ref:cleo}.
The first and second uncertainties are statistical and
systematic, respectively.
}
\label{table:summary}
\begin{ruledtabular}
\begin{tabular}{crr}
Measurement & Belle & Ref.~\cite{ref:cleo} 
\\
\hline
${\mathcal{B}(D^+ \rightarrow K^+ \eta )}$
& (1.08$\pm$0.17$\pm$0.08)$\times 10^{-4}$  &
$<$ 1.3$\times 10^{-4}$
\\ 
${\mathcal{B}(D^+\rightarrow K^+ \eta^\prime)}$ 
& (1.76$\pm$0.22$\pm$0.12)$\times 10^{-4}$&
$<$ 1.9$\times 10^{-4}$
\\
\end{tabular}
\end{ruledtabular}
\end{table}

%
%
 For our $A_{CP}$ measurement in 
the $D^+ \rightarrow \pi^+ \eta^{(\prime)}$ modes,
we re-optimize
our selection by maximizing $\mathcal{N}_S/\sigma_{S}$ where $\sigma_S$ is
the statistical error on the signal yield $\mathcal{N}_S$ in the simulated
sample. 
The re-optimized requirements for $D^+ \rightarrow \pi^+ \eta$ decays
are: 
$\xi$ $<$ $5^\circ$,
$\chi^2_\textrm{iso}$ $>$ 5,
$p_\eta$ $>$ 1.0 GeV/$c$, 
and 
$p^*_{D^{+}}$ $>$ 2.5 GeV/$c$, and for $D^+ \rightarrow \pi^+ \eta'$ 
are:
$\xi$ $<$ $5^\circ$,
$\chi^2_\textrm{iso}$ $>$ 2,
$p_{\eta^\prime}$ $>$ 1.0 GeV/$c$, 
and 
$p^*_{D^{+}}$ $>$ 2.5 GeV/$c$, respectively. These requirements are 
slightly less stringent than the selection criteria used for the
branching fraction measurements of DCS modes. 
This improves the statistical sensitivity on $A_{CP}$ by around 
15\%.

We determine the quantities 
$A^{D^{+}\rightarrow \pi^+ \eta^{(\prime)}}_{CP}$~\cite{ref:acp}
by measuring the asymmetry in signal yield
\begin{eqnarray}
A^{D^{+}\rightarrow \pi^+ \eta^{(\prime)}}_{\textrm{rec}}
&\equiv&
\frac
{N_\textrm{rec}^{D^{+}\rightarrow \pi^+ \eta^{(\prime)}}-N_\textrm{rec}^{D^{-}\rightarrow \pi^- \eta^{(\prime)}}}
{N_\textrm{rec}^{D^{+}\rightarrow \pi^+ \eta^{(\prime)}}+N_\textrm{rec}^{D^{-}\rightarrow \pi^- \eta^{(\prime)}}}
\nonumber \\
&\cong&
A^{D^{+}\rightarrow \pi^+ \eta^{(\prime)}}_{CP}
+
A^{D^{+}}_{FB} 
+
A^{\pi^{+}}_\epsilon,
\label{eq:arec}
\end{eqnarray}
where $N_\textrm{rec}$ is the number of reconstructed decays.
Note that we neglect the terms involving the product of asymmetries
and the 
approximation is valid for small asymmetries.
The measured asymmetry 
in Eq.~(\ref{eq:arec}) includes two contributions other than $A_{CP}$. 
One is the forward-backward asymmetry ($A^{D^{+}}_{FB}$)
due to $\gamma^*-Z^0$ interference in $e^+e^- \rightarrow c\bar{c}$
and the other is the detection efficiency asymmetry between positively
and negatively charged pions ($A^{\pi^{+}}_\epsilon$). 
To correct for the asymmetries
other than $A_{CP}$, we use a sample of 
Cabibbo-favored $D^+_s \rightarrow \phi \pi^+$ decays, in which  
the expected $CP$ asymmetry from the SM is negligible. Assuming that $A_{FB}$ is
the same for all charmed mesons, the difference between 
$A^{D^{+}\rightarrow \pi^+ \eta^{(\prime)}}_{\textrm{rec}}$ and
$A^{D^{+}_s\rightarrow \phi \pi^+}_{\textrm{rec}}$ yields the
$CP$ violation asymmetry 
$A^{D^{+}\rightarrow \pi^+ \eta^{(\prime)}}_{CP}$.
We reconstruct $\phi$ mesons
via the $K^+K^-$ decay channel,
requiring the $K^+K^-$ invariant mass to be between 1.01 and 1.03
GeV/$c^2$. This is the same technique as the one
developed in Ref.~\cite{ref:acp_kh}. 

 In order to obtain $A_{CP}$, we subtract the measured asymmetry
for $D^+_s \rightarrow \phi \pi^+$ from that for 
$D^{+}\rightarrow \pi^+ \eta^{(\prime)}$ in three-dimensional (3D) 
bins, where the 3D bins are  
the transverse momentum, $p^\textrm{lab}_{T\pi}$, 
and the polar angle of the $\pi^+$ in the laboratory system,
$\cos{\theta}^\textrm{lab}_\pi$, and the charmed meson
polar angle in the center-of-mass system,
$\cos{\theta}^*_{D^+_{(s)}}$. 
Simultaneous fits to the $D^+_{(s)}$ and $D^-_{(s)}$ invariant mass 
distributions for each bin are carried out.
A double Gaussian for the signal and a linear function for the
background are used as PDFs for $D^+_s \rightarrow \phi \pi^+$.
The average value over all bins is found to be 
$A_\textrm{rec}^{D^+_s \rightarrow \phi \pi^+}$=
(0.17 $\pm$ 0.13)\%. 
After the subtraction of $A_\textrm{rec}^{D^+_s \rightarrow \phi \pi^+}$
component, weighted averages
of the $A_{CP}$ values summed over the 3D bins are
($+$1.74 $\pm$ 1.14)\% and  
($-$0.12 $\pm$ 1.13)\% 
for $D^+ \rightarrow \pi^+ \eta$ and 
$D^+ \rightarrow \pi^+ \eta^{\prime}$, respectively, 
where the uncertainties originate from the finite size of
the $D^+ \rightarrow \pi^+ \eta$ (1.13\%),
$D^+ \rightarrow \pi^+ \eta'$ (1.12\%),
and
$D^+_s \rightarrow \phi \pi^+$ (0.13\%) samples.
The $\chi^2/$d.o.f values summed over the 3D bins  
are 
28.7/11=2.6
for $D^+ \rightarrow \pi^+ \eta$ 
and
15.7/11=1.4 
for $D^+ \rightarrow \pi^+ \eta'$.

 The dominant source of systematic uncertainty in the $A_{CP}$ 
measurement is the uncertainty in the $A^{D^+_s \rightarrow \phi \pi^+}_\textrm{rec}$
determination, which originates from the following sources: the statistics
of the 
$D^+_s \rightarrow \phi \pi^+$ sample (0.13\%),
possible detection asymmetry of kaons from 
$\phi \rightarrow K^+K^-$ (0.05\%)~\cite{ref:marko}
and the choice of binning for the 3D map (0.12\%, 0.01\%), for  
$D^+ \rightarrow \pi^+ \eta$ 
and
$D^+ \rightarrow \pi^+ \eta'$,
respectively. Another source is the fitting of the invariant mass distribution
(fit interval, choice of the fitting function), which 
contributes uncertainties of
0.05\% to $A^{D^+ \rightarrow \pi^+ \eta}_{CP}$, 
and 
0.07\% to $A^{D^+ \rightarrow \pi^+ \eta'}_{CP}$.
Possible systematic uncertainties due to the fixed signal PDF parameters are
estimated to be 0.01\% for  
$A^{D^+ \rightarrow \pi^+ \eta}_{CP}$
and 
0.07\% for
$A^{D^+ \rightarrow \pi^+ \eta'}_{CP}$.
By
combining all sources in quadrature, we obtain
$A_{CP}^{D^+ \rightarrow \pi^+ \eta}$
=
($+$1.74 $\pm$ 1.13 $\pm$ 0.19)\% and  
$A_{CP}^{D^+ \rightarrow \pi^+ \eta'}$
=
($-$0.12 $\pm$ 1.12 $\pm$ 0.17)\%. These are
the most precise measurements of $A^{D^+ \rightarrow \pi^+ \eta^{(\prime)}}_{CP}$ to date.


 In conclusion, we report the first observation of 
DCS $D^+ \rightarrow K^+ \eta^{(\prime)}$ decays
using a 791 fb$^{-1}$ data sample collected with the Belle detector
at the KEKB asymmetric-energy $e^+e^-$ collider.
The ratios of branching fractions of DCS modes 
with respect to the SCS modes are
$\mathcal{B}(D^+ \rightarrow K^+ \eta)$/$\mathcal{B}(D^+ \rightarrow \pi^+ \eta)$
= (3.06 $\pm$ 0.43 $\pm$ 0.14)\%
and
$\mathcal{B}(D^+ \rightarrow K^+ \eta')$/$\mathcal{B}(D^+ \rightarrow \pi^+ \eta')$
= (3.77 $\pm$ 0.39 $\pm$ 0.10)\%.
Using our DCS branching fractions and that of $D^0 \rightarrow K^+ \pi^0$ from
Ref.~\cite{ref:cleo}, the first measurement of the relative phase
difference between the 
tree and annihilation amplitudes in $D^+$ decays
is reported with $\delta_{TA}$ = (72 $\pm$ 9)$^\circ$ or
(288 $\pm$ 9)$^\circ$ using the technique suggested
in Ref.~\cite{ref:dcsd_th};
this is important information relevant to final-state interactions.
We also search for $CP$ asymmetries in SCS modes down to the $\mathcal{O}$(\%)
level. 

We thank the KEKB group for excellent operation of the
accelerator, the KEK cryogenics group for efficient solenoid
operations, and the KEK computer group and
the NII for valuable computing and SINET4 network support.  
We acknowledge support from MEXT, JSPS and Nagoya's TLPRC (Japan);
ARC and DIISR (Australia); NSFC (China); MSMT (Czechia);
DST (India); MEST, NRF, NSDC of KISTI, and WCU (Korea); MNiSW (Poland); 
MES and RFAAE (Russia); ARRS (Slovenia); SNSF (Switzerland); 
NSC and MOE (Taiwan); and DOE (USA). E. Won 
acknowledges support by NRF Grant No. 2011-0027652 and B. R. Ko
acknowledges support by NRF Grant No. 2011-0025750.

\end{document}